\documentclass[seceq]{ptptex}

\usepackage{graphicx}

\title{
Galaxy N-z Relation and CMB Spectrum Based on Cosmological Model with Scalar Field Having Modified Potential Form
}

\author{
Koichi \textsc{
Hirano}%
, 
Kiyoshi \textsc{
Kawabata}%
, 
Zen \textsc{
Komiya}%
, and 
Hiroshi \textsc{
Bunya}%
}

\inst{
Department of Physics, Tokyo University of Science, Shinjuku-ku, Tokyo, Japan
}

\abst{
We have succeeded in establishing a cosmological model with a non-minimally coupled scalar field $\phi$ that can account not only for the spatial periodicity or the {\it picket-fence structure} exhibited by the galaxy $N$-$z$ relation of the 2dF survey, but also for the spatial power spectrum of the cosmic microwave background radiation (CMB) temperature anisotropy observed by the WMAP satellite.
The scalar field of our model universe starts from an extremely small value at around the nucleosynthesis epoch,
remains in that state for sufficiently long periods, allowing sufficient time for the CMB temperature anisotropy to form, and then starts to grow in magnitude at the redshift $z$ of $\sim 1$, followed by a damping oscillation which is required 
to reproduce the observed picket-fence structure of the $N$-$z$ relation. 
To realize such behavior of the scalar field, we have found it necessary to introduce a new form of potential $V(\phi)\propto \phi^2\exp(-q\phi^2)$, with $q$ being a constant. Through this parameter $q$, we can control the epoch at which the scalar field starts growing.
}

\begin{document}

\maketitle

\section{Introduction}
The $N$-$z$ relation obtained from the 2dF Galaxy Redshift Survey (2dF GRS) exhibits a definite picket-fence structure at the interval of roughly $\Delta z=0.03$ (see, e.g., Fig. 17 of Colless et al. 2001). The aim of this paper is to establish a cosmological model with a non-minimally coupled scalar field $\phi$ that can account for both the spatial periodicity or the {\it picket-fence structure} exhibited by the galaxy $N$-$z$ relation of the 2dF survey and the spatial power spectrum of the CMB temperature anisotropy observed by the WMAP satellite. \cite{hin07,spe07}

\section{Model and Observational Tests}
The relevant Lagrangian $L$ for our cosmological model is
\begin{equation}
L=\frac{1}{2}\xi R\psi^2-\frac{c^4}{16\pi G}R+\frac{1}{2}g^{\mu\nu}\partial_{\mu}\psi\partial_{\nu}\psi-\frac{1}{2}\left(\frac{m_\phi c}{\hbar}\right)^2\psi^2\exp{\left(-q\frac{4\pi G}{3c^4}\psi^2\right)}+\frac{c^4}{8\pi G}\mit\Lambda +L^{(\rm mr)}. \label{Lagrangean}
\end{equation}
The time evolution equations for the cosmic scale factor $a$, and the scalar field $\phi$ for our cosmological model are as shown below:
\begin{eqnarray}
\frac{\ddot{a}}{a} & = & \left[2\frac{\dot{a}^2}{a^2}(1-6\xi\phi^2)-3(\dot{\phi})^2-24\xi\frac{\dot{a}}{a}\phi\dot{\phi}+6\xi(\dot{\phi})^2-6\xi a^2\left(\frac{m_\phi c}{\hbar}\right)^2(1-q\phi^2)\phi^2e^{(-q\phi^2)}\right. \nonumber \\
 & & \left.-\frac{3}{2}\frac{H_0^2}{c^2}\left(\frac{\Omega_{m,0}}{a}+\frac{\Omega_{r,0}}{a^2}\right)\right]\left(1-6\xi\phi^2(1-6\xi)\right)^{-1},   \label{EqAdots}
\end{eqnarray}
\begin{equation}
\hspace{-47mm}
\ddot{\phi}=-2\frac{\dot{a}}{a}\dot{\phi}-6\xi\frac{\ddot{a}}{a}\phi-a^2\left(\frac{m_\phi c}{\hbar}\right)^2(1-q\phi^2)\phi\exp{(-q\phi^2)}. \label{EqPhidots}
\end{equation}
Note that the potential term (the fourth term on the right-hand side of Eq.(\ref{Lagrangean})) reflects our modification to the original form employed by Morikawa(1991), which coincides with the latter if $q=0$.
This new form of potential enables us to control the epoch when the growth of the scalar field $\phi$ begins to take place.(Fig.1) The closer this epoch is to the present era, the less affected are the amplitudes of the spatial power spectrum of the CMB temperature anisotropy in the large-scale domain because of the reduced  effect  of  the late-time integrated Sachs-Wolfe effect.
\begin{figure}
\begin{minipage}{1.00\hsize}
\begin{center}
\includegraphics[width=84mm]{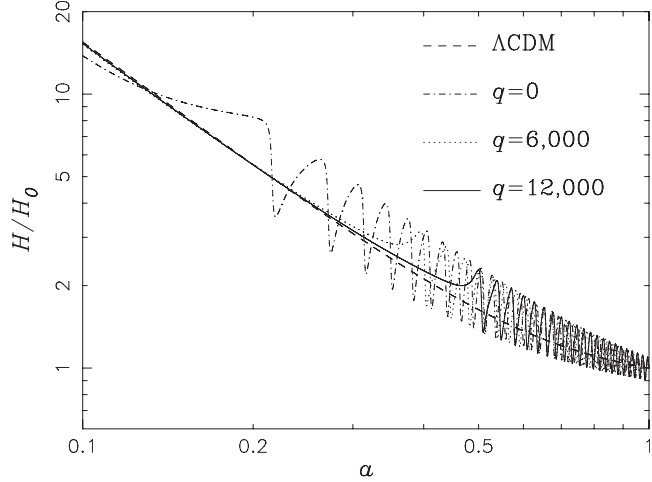}
   \caption{Dependence on $q$ of the onset time of the oscillation of the Hubble parameter $H/H_0$ normalized to the Hubble constant. The abscissa is the scale factor $a$.}
   \label{fig:1}
\end{center}
\end{minipage}
\end{figure}
\begin{figure}
\begin{minipage}{1.00\hsize}
\begin{center}
       \includegraphics[width=84mm]{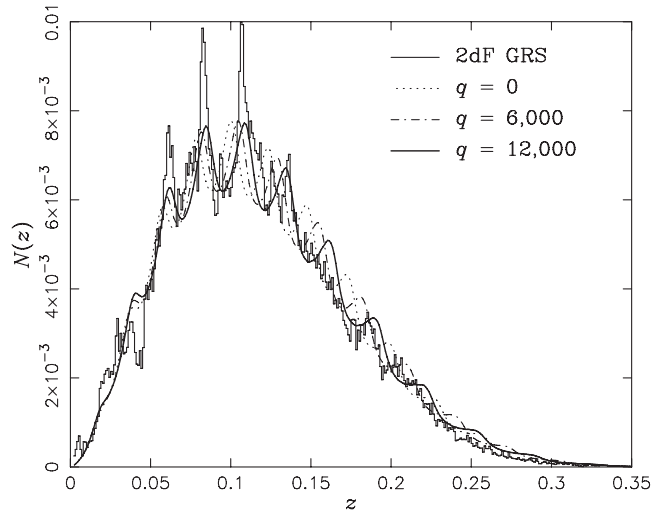}
   \caption{Dependence on  $q$ of the computed $N$-$z$ relationship of galaxies. The abscissa indicates the redshift $z$, while the ordinate is the fractional number count $N$.}
   \label{fig:2}
\end{center}
\end{minipage}
\end{figure}
\begin{figure}
\begin{minipage}{1.00\hsize}
\begin{center}
       \includegraphics[width=84mm]{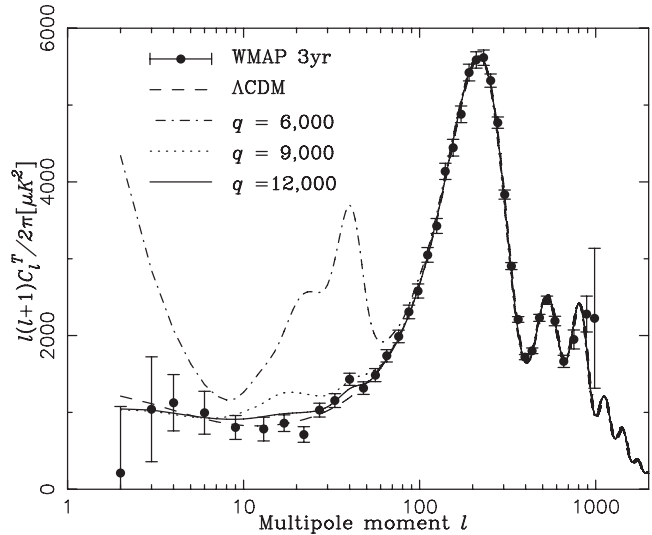}
   \caption{Spatial power spectra of the CMB temperature anisotropy computed for our scalar field models with $q=$6,000, 9,000, and 12,000, respectively.
}
   \label{fig:3}
\end{center}
\end{minipage}
\end{figure}
\par
Our model (q=12000) having an oscillating expansion rate can account well for both the spatial periodic structure of the galaxy $N$-$z$ relation of the 2dF survey(Fig.2) and the spatial power spectrum of CMB anisotropy observed by the WMAP(Fig.3).

\section{Conclusion}
We have succeeded in establishing a cosmological model with a non-minimally coupled scalar field $\phi$ that can account not only for the spatial periodicity or the {\it picket-fence structure} exhibited by the galaxy $N$-$z$ relation of the 2dF survey but also for the spatial power spectrum of the CMB temperature anisotropy observed by the WMAP satellite.
The parameter values of the proposed model are as follows:
$\Omega_{\rm b,0}h_{100}^2=0.024$, $\Omega_{\rm m,0}=0.237$, $H_0=72$ km ${\rm s^{-1} Mpc^{-1}}$, $\Omega_{\rm \phi,0}=0.15$, $\Omega_{{\mit \Lambda},0}=0.613$, $\xi=-40$, $m_\phi=3.7\times 10^{-31}h_{100}$~eV, $q=12000$,
$\tau=0.089$, $n_{\rm s}=1.0$, yielding the cosmic age of 13.2 Gyrs, where $h_{100}\equiv H_0/100$, $\xi$ is the curvature scalar coupling constant, $m_\phi$ the mass of the scalar field, $n_{\rm s}$ the scalar spectral index, and $\tau$ the optical depth of the universe.


%


\begin{thebibliography}{99}
 
\bibitem[1]{col01} Colless, M. M., et al., Mon. Not. R. Astron. Soc., \textbf{328}, (2001), 1039.
\bibitem[2]{hin07} Hinshaw, G., et al., Astrophys. J. Suppl. Ser., \textbf{170}, (2007), 288.
\bibitem[3]{spe07} Spergel, D. N., et al., Astrophys. J. Suppl. Ser., \textbf{170}, (2007), 377.
\bibitem[4]{mor91} Morikawa, M., Astrophys. J., \textbf{369}, (1991), 20.
\end{thebibliography}
\end{document}